\documentclass{article}

\usepackage{xspace}
\usepackage{graphicx}
\usepackage{cite}
\usepackage{amsthm}
\usepackage{amsmath,amssymb}
\usepackage{url}

\theoremstyle{theorem}
\newtheorem{theorem}{Theorem}[section]
\newtheorem{cor}[theorem]{Corollary}
\newtheorem{lemma}[theorem]{Lemma}
\newtheorem{prop}[theorem]{Proposition}

\newtheorem{conj}[theorem]{Conjecture}

\theoremstyle{definition}
\newtheorem{definition}[theorem]{Definition}

\theoremstyle{remark}
\newtheorem{remark}[theorem]{Remark}

\newtheorem{obs}[theorem]{Observation}

\newcommand{\ie}{\emph{i.e.}}

\newcommand{\eg}{\emph{e.g.}}
\newcommand{\etc}{\emph{etc}}
\newcommand{\etal}{\emph{et al}}

\newcommand{\PAR}{\ensuremath{\mathsf{PAR}}}

\newcommand{\disjoint}{{\sc Dis\-joint\-ness}\ensuremath{_k}}
\newcommand{\intersection}{{\sc In\-ter\-sec\-tion}\ensuremath{_k}}

\newcommand{\svn}{\ensuremath{v^0}}
\newcommand{\svy}{\ensuremath{v^1}}
\newcommand{\shn}{\ensuremath{h^0}}
\newcommand{\shy}{\ensuremath{h^1}}

\newcommand{\vy}{\ensuremath{v^1}}
\newcommand{\vn}{\ensuremath{v^0}}
\newcommand{\nv}{\ensuremath{nV^0}}
\newcommand{\hy}{\ensuremath{h^1}}
\newcommand{\hn}{\ensuremath{h^0}}
\newcommand{\nh}{\ensuremath{nH^0}}
\newcommand{\paratio}{\ensuremath{\mathsf{PAR}}}

\begin{document}

\title{Approximate Privacy: PARs for Set Problems}

\author{Joan Feigenbaum\thanks{Supported in part by NSF grants
0331548 and 0716223 and IARPA grant FA8750-07-0031.}\\
Department of Computer Science\\
Yale University\\
\texttt{joan.feigenbaum@yale.edu}
\and
Aaron D.\ Jaggard\thanks{Supported in part by
NSF grants 0751674 and 0753492.}\\
DIMACS\\
Rutgers University\\
\texttt{adj@dimacs.rutgers.edu}
\and Michael Schapira\thanks{Supported by NSF grant 0331548.}\\
Department of Computer Science\\
Yale University\\
and\\
Computer Science Division\\
University of California, Berkeley\\
\texttt{michael.schapira@yale.edu}
}

\date{}

\maketitle

\begin{abstract}
In previous work (arXiv:0910.5714), we introduced the Privacy Approximation Ratio (PAR) and used it to study the privacy of protocols for second-price Vickrey auctions and Yao's millionaires problem.  Here, we study the PARs of multiple protocols for both the disjointness problem (in which two participants, each with a private subset of $\{1,\ldots,k\}$, determine whether their sets are disjoint) and the intersection problem (in which the two participants, each with a private subset of $\{1,\ldots,k\}$, determine the intersection of their private sets).

We show that the privacy, as measured by the PAR, provided by any protocol for each of these problems is necessarily exponential (in $k$).  We also consider the ratio between the subjective PARs with respect to each player in order to show that one protocol for each of these problems is significantly fairer than the others (in the sense that it has a similarly bad effect on the privacy of both players).
\end{abstract}

\newpage

\section{Introduction}\label{sec-intro}

Widespread use of computers and networks in almost all aspects of daily life has led to a
proliferation of sensitive electronic data records and thence to extensive study of privacy-preserving
computation.  One fruitful approach is based on the combinatorial characterization
of privately computable functions put forth by Chor and Kushilevitz~\cite{CK91} and the subsequent
communication-complexity analysis of privately computable functions by Kushilevitz~\cite{K92}.  Using this
approach, one can show, for example, that Yao's millionaires' problem~\cite{Y79} is not
perfectly privately computable~\cite{CK91} and that the two-bidder, $2^{nd}$-price Vickrey
auction is perfectly privately computable but only at the cost of and exponential amount of
communication by the bidders~\cite{BS}.

Motivated by the fact that functions of interest may not be perfectly
privately computable or may be so only by impractically costly protocols, we began
in \cite{fjs09tr14} a communication-complexity-based investigation of approximate privacy.
We formulated both worst-case and
average-case versions of the {\it privacy-approximation ratio} (PAR) of a
function $f$ in order to quantify the {\it amount} of privacy that can be
preserved by a protocol that computes $f$ and studied the tradeoff between
approximate privacy and communication complexity in protocols for the millionaires' problem
and the two-bidder, $2^{nd}$-price Vickrey auction.

Informally, a two-party protocol is \emph{perfectly
privacy-preserving} if the two parties (or a third party observing
the communication between them) cannot learn more from the execution
of the protocol than the value of the function the protocol
computes.  (This notion can be extended naturally to protocols
involving more than two participants, but we do not consider the more general notion
in this paper.)  Chor and Kushilevitz~\cite{CK91,K92} formalize this notion of privacy using
the communication-complexity-theoretic notions of the {\it ideal monochromatic regions} of
a function $f$ and the {\it monochromatic rectangles} of a protocol $P$ that computes $f$.
Every two-input function $f$ can be represented by a two-dimensional matrix $A(f)$ in which
$A(f)_{(x_1,x_2)} = f(x_1,x_2)$.  In the partition of $A(f)$ into the ideal monochromatic
regions of $f$, the entries $A(f)_{(x_1,x_2)}$ and $A(f)_{(y_1,y_2)}$ are in the same region
if and only if $f(x_1,x_2)=f(y_1,y_2)$; if $f$ is perfectly privately computable, then there
is a protocol $P$ for $f$ that partitions $A(f)$ into a set of monochromatic rectangles that
is exactly equal to the set of ideal monochromatic regions of $f$.  For functions that are
not perfectly privately computable, our notions of approximate privacy~\cite{fjs09tr14}
quantify the worst-case and average-case ratios between the size of an ideal monochromatic
region of $f$ and the corresponding monochromatic rectangle in the partition induced by a
maximally privacy-preserving protocol for $f$.

In this paper, we apply our PAR framework to the intersection problem (in which party 1's input
is a set $S_1$, party 2's input is a set $S_2$, and the goal of the protocol is to compute
$S_1 \cap S_2$) and to its decision version disjointness (in which $f(S_1, S_2) = 1$ if
$S_1 \cap S_2 = \emptyset$, and $f(S_1, S_2) = 0$ otherwise).  From both the privacy perspective
and the communication-complexity perspective, these are extremely natural problems to study.
The intersection problem has served as a motivating example in the study of privacy-preserving
computation for decades; in a typical application, two organizations wish to compute the set of
members that they have in common without disclosing to each other the people who are members of
only one of the organizations.  The disjointness problem plays a central role in the theory and
application of communication complexity, where the fact that $n+1$ bits of communication are
required to test disjointness of two subsets of $\{1, \ldots, n\}$ is used to prove many
worst-case lower bounds.

\subsection{Our Findings}

In applying our PAR framework to the disjointness and intersection problems, we consider three natural
protocols that apply to both problems.  We compute the objective and subjective PARs for all three protocols for both problems.  The objective and subjective PARs are
exponential in all cases, but we show that the protocol that is intuitively the
best is quantifiably (and significantly) more fair than the others in the sense
described below; to do this, we consider the ratios of the subjective PARs (as described in Sec.~\ref{ssec:ratio}) and argue that this captures some intuitive sense of fairness.  Table~\ref{tab:results} in Sec.~\ref{sec:overview} summarizes our
results for PAR values for the various problems and protocols that we consider here; the corresponding theorems and proofs are in Secs.~\ref{sec:disjoint} and~\ref{sec:intersection}.

\subsection{Related Work: Defining Privacy-Preserving
Computation}

In addition to Brandt and Sandholm~\cite{BS}, who used Kushilevitz's formulation of privacy-preserving
computation to prove an exponential lower bound on the communication complexity of
privacy-preserving $2^{nd}$-price Vickrey auctions, the privacy work of Bar-Yehuda {\it et
al.}~\cite{BCKO} is also based on the communication-complexity framework of \cite{CK91,K92}.

Among other approaches to privacy-preserving computation, the most extensively developed
is that of {\it secure, multiparty computation} (SMC).  As observed by Brandt and Sandholm~\cite{BS},
bidders' privacy in online auctions, which was our original motivation as well as theirs, could in
principle be achieved by starting with a strategyproof mechanism and then having the agents
themselves compute the outcome and payments using an SMC protocol.  This approach has been followed
successfully by, for example, Dodis, Halevi, and Rabin~\cite{DHR00} and Naor, Pinkas, and
Sumner~\cite{NPS99} but, as discussed in more detail \cite{BS,fjs09tr14}, can in general require
assumptions about the strategic nature of the computational nodes that do not apply to bidders in
auctions, unproven cryptographic assumptions, or excessive communication costs.  Thus, non-SMC
approaches are worth pursuing.

In our study of PAR, we consider protocols that compute exact results but
preserve privacy only approximately.  Several works, including \cite{BCNW,FIMNSW,HKKN}, have considered
protocols that compute approximate results in a privacy-preserving manner, but they are unrelated to the
questions we ask here.  Similarly, definitions and techniques from {\it differential
privacy}~\cite{DiffPrivSurvey} (and its mechanism-design extensions~\cite{mt07focs,GRS09}) are aimed at
computing approximate results and are inapplicable to the problems that we study here.

\subsection{Paper Outline}

In Sec.~\ref{sec:approx}, we review the PAR framework of~\cite{fjs09tr14} and discuss the ratios of average-case subjective PARs.  Section~\ref{sec:overview} gives formal definitions of the problems we study, describes the protocols for these problems that we consider, and gives a summary and discussion of our PAR results.  Sections~\ref{sec:disjoint} and~\ref{sec:intersection} give the full statements and proofs of our PAR results.  Section~\ref{sec:conc} discusses avenues for future work.  Appendix~\ref{ap:model} provides additional background about our approach, and App.~\ref{ap:alternate}.  Sections~\ref{ssec:par-wc} and \ref{ssec:par-ac} and Apps.~\ref{ap:model} and~\ref{ap:alternate} are drawn from~\cite{fjs09tr14}; we include them here for the convenience of the reader.

\section{Privacy Approximation Ratios}\label{section-PAR-definitions}\label{sec:approx}

We now review our formulations of Privacy Approximation Ratios (PARs)~\cite{fjs09tr14}.
We refer readers to Section~\ref{subsec_perfect-privacy} of the Appendix below for a more
thorough explanation. We assume that the reader is familiar with Yao's model of two-party
communication. Readers unfamiliar with this material should refer to Section~\ref{subsec-Yao}
of the Appendix below or, for a more in-depth treatment, to Kushilevitz and Nisan~\cite{KN97}.

Chor and Kushilevitz~\cite{CK91,K92} put forth definitions and characterizations of perfectly
private communication protocols. Their framework was further developed in \cite{fjs09tr14},
where we introduced the notion of PARs. In this paper, as in \cite{fjs09tr14}, we deal only
with \emph{deterministic} communication protocols, but the framework can be extended to randomized
protocols.

As explained in the previous section, there are natural problems for which
perfect privacy is either impossible or very costly (in terms of communication complexity) to obtain.
\emph{Privacy-approximation ratios} (PARs) allow us to quantify how
well a protocol preserves privacy relative to the ideal (but perhaps impossible to implement)
computation of the outcome of a problem.  Approximate privacy has both worst-case and average-case
formulations.

\subsection{Worst-Case PARs}\label{ssec:par-wc}

Any function $f:\{0,1\}^k\times \{0,1\}^k\rightarrow\{0,1\}^t$ can
be visualized as a $2^k\times 2^k$ matrix with entries in
$\{0,1\}^t$, in which the rows represent the possible inputs of
party $1$, the columns represent the possible inputs of party $2$,
and each entry contains the value of $f$ associated with its row and
column inputs. This matrix is denoted $A(f)$.

For any communication protocol $P$ for a function $f$, let
$R^P(x_1,x_2)$ denote the monochromatic rectangle in $A(f)$ induced by $P$ for the pair
of inputs $(x_1,x_2)$. Let $R^I(x_1,x_2)$ denote the maximal monochromatic region in $A(f)$
containing $A(f)_{(x_1,x_2)}$, \emph{i.e.}, the maximal set of entries in $A(f)$ that contain
the value $f(x_1,x_2)$. Intuitively, $R^P(x_1,x_2)$ is the set
of inputs that are indistinguishable from $(x_1,x_2)$ to this particular protocol $P$.
Similarly, $R^I(x_1,x_2)$ is the set
of inputs that \emph{would be} indistinguishable from $(x_1,x_2)$ to a
perfectly private protocol if such a protocol existed.
We wish to quantify how far $P$ is from a hypothetical ideal protocol in
terms of indistinguishability of inputs. Let $|R|$ denote the \emph{size} or
cardinality of $R$, \ie, the number of inputs in $R$.

\begin{definition}[Worst-case objective PAR of $P$]
The \emph{worst-case objective privacy-ap\-prox\-i\-ma\-tion ratio} of
communication protocol $P$ for function $f$ is
\[
    \alpha=\max_{(x_1,x_2)}\ \frac{|R^I(x_1,x_2)|}{|R^P(x_1,x_2)|}.
\]

We say that $P$ is \emph{$\alpha$-objective-privacy-preserving in
the worst case}.
\end{definition}

Given any region $R$ in the matrix $A(f)$, if party
1's private input is $x$, then party 1 can use this
knowledge to eliminate all entries in $R$ outside of row $x$; similarly,
party 2 can eliminate all parts of $R$ outside of the appropriate column.
Hence, the other parties should be concerned not with all of $R$ but rather
with what we call the \emph{$i$-partitions} of $R$.

\begin{definition} [$i$-partitions]
The \emph{$1$-partition} of a region $R$ in a matrix $A$ is the set of
disjoint rectangles $R_{x_1}=\{x_1\}\times\{x_2\ s.t.\
(x_1,x_2)\in R\}$ (over all possible inputs $x_1$).
\emph{$2$-partitions} are defined analogously.
\end{definition}

\begin{definition} [$i$-induced tilings]
The \emph{$i$-induced tiling} of a protocol $P$ is the refinement of the
tiling induced by $P$ obtained by $i$-partitioning each rectangle in
it.
\end{definition}

\begin{definition} [$i$-ideal monochromatic partitions]
The \emph{$i$-ideal monochromatic partition} is the refinement of the ideal
monochromatic partition obtained by $i$-partitioning each region in
it.\end{definition}

If $P$ is a communication protocol for the function $f$, then we let
$R_i^P(x_1,x_2)$ denote the monochromatic rectangle containing
$A(f)_{(x_1,x_2)}$ in the $i$-induced tiling for $P$. Similarly, we let
$R_i^I(x_1,x_2)$ denote the monochromatic rectangle containing
$A(f)_{(x_1,x_2)}$ in the $i$-ideal monochromatic partition of
$A(f)$.

\begin{definition}[Worst-case PAR of $P$ with respect to $i$]
\ The \emph{worst-case privacy-ap\-prox\-i\-ma\-tion ratio with respect to
$i$} of communication protocol $P$ for function $f$ is
\[
    \alpha=\max_{(x_1,x_2)}\ \frac{|R_i^I(x_1,x_2)|}{|R^P_i(x_1,x_2)|}.
\]

We say that $P$ is \emph{$\alpha$-privacy-preserving with respect to
$i$ in the worst case}.
\end{definition}

\begin{definition}[Worst-case subjective PAR of $P$]
The \emph{worst-case subjective privacy-ap\-prox\-i\-ma\-tion ratio} of
communication protocol $P$ for function $f$ is the maximum, over $i=1,2$, of the
worst-case privacy-approximation ratio with respect party $i$.
\end{definition}

\begin{definition}[Worst-case PAR]
The \emph{worst-case objective (subjective) PAR for a function $f$}
is the minimum, over all protocols $P$ for $f$, of the worst-case objective (subjective) PAR of $P$.
\end{definition}

\subsection{Average-Case PARs}\label{ssec:par-ac}

As we showed in \cite{fjs09tr14}, good approximate privacy may be
just as unobtainable as perfect privacy if one insists on worst-case
bounds. Thus, we also consider average-case PAR, {\it i.e.},
the \emph{average} ratio between the size of the monochromatic
rectangle containing the private inputs and the corresponding region
in the ideal monochromatic partition.

\begin{definition}[Average-case objective PAR of $P$]\label{def:par-avg}
Let $D$ be a probability distribution over the space of inputs. The
\emph{average-case objective privacy-approximation ratio} of
communication protocol $P$ for function $f$ is
\[
    \alpha = E_{D}\ [\frac{|R^I(x_1,x_2)|}{|R^P(x_1,x_2)|}].
\]

We say that $P$ is \emph{$\alpha$-objective privacy-preserving in
the average case with distribution $D$} (or \emph{with respect to
$D$}).
\end{definition}

We define average-case PAR with respect to $i$ analogously and average-case
subjective PAR as the maximum over $i$ of the average-case PAR with respect to player $i$.
Finally, we define the \emph{average-case objective (subjective) PAR for a function
$f$} as the minimum, over all protocols $P$ for $f$, of the average-case objective
(subjective) PAR of $P$.

In computing the average-case PAR (either objective or subjective) with respect
to the uniform distribution, we may simplify the previous expressions for PAR values.
If each player's value space has $k$ bits, then the average-case objective PAR with
respect to the uniform distribution equals
\begin{equation*}
\mathsf{PAR}_k = \sum_{(x_1,x_2)} \frac{1}{2^{2k}} \frac{|R^I(x_1,x_2)|}{|R^{P}(x_1,x_2)|},
\end{equation*}
where the sum is over all pairs $(x_1,x_2)$ in the value space.
We may combine all of the terms corresponding to points in the same protocol-induced
rectangle to obtain
\begin{equation}
\mathsf{PAR}_k = \sum_{S} \frac{|S|}{2^{2k}} \frac{|R^I(S)|}{|S|} = \frac{1}{2^{2k}} \sum_{S} |R^I(S)|,\label{eq:pargen}
\end{equation}
where the sums are now over protocol-induced rectangles $S$.  Note also that the
average-case PAR with respect to $i$ and with respect to the uniform distribution
is obtained by replacing $R^I(S)$ with $R^I_i(S)$ in Eq.~\ref{eq:pargen}.

It may seem that a probability-mass-based definition of average-case PAR should be used instead, \ie, that the occurrences of set cardinality in the quantity considered in Def.~\ref{def:par-avg} should be replaced by the probability measure of the regions in question.  However, as we discuss in~\cite{fjs09tr14}, such a definition is unable to distinguish between examples that should be viewed as having very different levels of privacy; by contrast, the definition that we consider here is able to distinguish between such cases.

\subsection{Ratios of Subjective PARs}\label{ssec:ratio}

Here we introduce a new quantity that we did not consider in~\cite{fjs09tr14}.  Given some protocol $P$ for a function $f$, let $\paratio^i_D(k)$ be the average-case subjective PAR of $P$ with respect to protocol participant $i$ and distribution $D$ on the $k$-bit input space.  We then let
\[
\paratio^\mathsf{max}_D(k) = \max_i\paratio^i_D(k) \qquad\text{ and }\qquad\paratio^\mathsf{min}_D(k) = \min_i\paratio^i_D(k),
\]
where the $\max$ and $\min$ are taken over all protocol participants.  We then define the \emph{ratio of (average-case) subjective PARs} to be
\[
\frac{\paratio^\mathsf{max}_D(k)}{\paratio^\mathsf{min}_D(k)}\geq 1.
\]
Intuitively, in a two-participant protocol, this captures how much greater a negative effect the protocol $P$ can have on one participant than on the other participant. The average-case subjective PAR of a protocol $P$ identifies the maximum effect that $P$ can have on the privacy with respect to a participant.  However, it does not capture whether this effect is similar for both players, and in fact this effect can be quite different.  Below we show that, for both the disjointness and intersection problems, there are protocols that have exponentially large subjective PARs; for some protocols, the subjective PAR with respect to one player is exponentially larger than that with respect to the other player, while for one protocol for each problem, the subjective PARs with respect to the different players differ only by a constant (asymptotic) factor.  We argue that this is an important distinction and that the ratio of average-case subjective PARs captures some intuitive notion of the fairness of the protocol.  If a protocol has a much larger PAR with respect to player $2$ than with respect to player $1$, an agent might agree to participate in a protocol run only if he is assigned the role of player $2$ (so that he learns much more about the other player than the other player learns about him).  Thus, from the perspective of the protocol implementer who needs to induce participation, protocols with small ratios of average-case subjective PARs would likely be more desirable.

\section{Overview of Problems, Results, and Protocols}\label{sec:overview}

We now provide an overview of our PAR results and discuss their significance.  We start with technical definitions of the problems and protocols that we consider here.

\subsection{Problems}

We define the \disjoint\ problem as follows:

\noindent\textbf{Problem:} \disjoint

\noindent\textbf{Input:} Sets $S_1,S_2\subseteq\{1,\ldots,k\}$ encoded by $x_1$ and $x_2$.

\noindent\textbf{Output:} $1$ if $S_1\cap S_2 = \emptyset$, $0$ if $S_1\cap S_2 \neq \emptyset$.

\begin{figure}[htp]
\begin{center}
\includegraphics[height=2in]{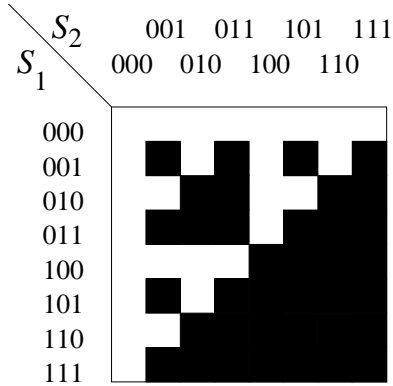}
\caption{Ideal monochromatic partition for \disjoint\ with $k=3$.}\label{fig:disk3}
\end{center}
\end{figure}

Figure~\ref{fig:disk3} illustrates the ideal monochromatic partition of the $3$-bit value space; inputs for which $S_1$ and $S_2$ are disjoint are white, and inputs for which these sets are not disjoint are black.

We define the \intersection\ problem as follows:

\noindent\textbf{Problem:} \intersection

\noindent\textbf{Input:} Sets $S_1,S_2\subseteq\{1,\ldots,k\}$.

\noindent\textbf{Output:} The set $S_1\cap S_2$.

\begin{figure}[htp]
\begin{center}
\includegraphics[height=2in]{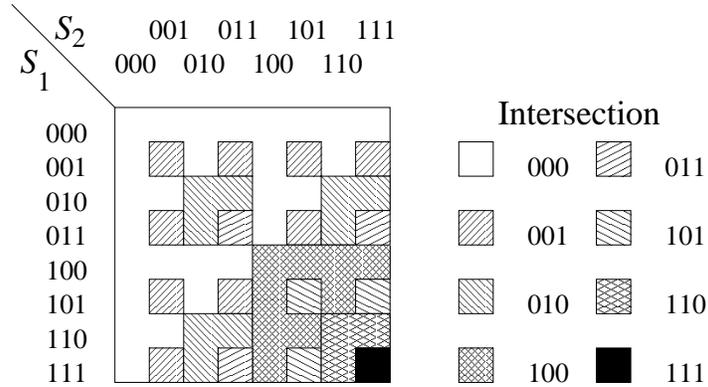}
\end{center}
\caption{Ideal monochromatic partition for \intersection\ problem with $k=3$.}\label{fig:intk3}
\end{figure}

Figure~\ref{fig:intk3} shows the ideal monochromatic partition of the $3$-bit value space for \intersection.  The key at the right indicates the output set.  (Here, as throughout this paper, we encode $S\subseteq \{1,\ldots,k\}$ as bitstring of length $k$ in which the most significant bit is $1$ if $k\in S$, \etc., so that $1011$ encodes $\{1,2,4\}\subset\{1,2,3,4\}$; we will abuse notation and identify $x\in\{0,1\}^k$ with the subset of $\{1,\ldots,k\}$ that it encodes.)

\subsection{Protocols}

For each problem, we identify three possible protocols for computing the output of the problem.  We describe these protocols here; in Secs.~\ref{sec:disjoint} and~\ref{sec:intersection} we discuss the structure of the tilings that these protocols induce and illustrate these tilings for $k=1,2,3$.

\paragraph{Trivial protocol} In the trivial protocol, player $1$ (w.l.o.g.) sends his input to player $2$, who determines computes the output and sends this back to player $1$.  This requires the transmission of $k+1$ bits for \disjoint\ and $2k$ bits for \intersection.

\paragraph{$1$-first protocol} In the $1$-first protocol, player $1$ announces a bit, and player $2$ replies with his corresponding bit if its value might affect the output (\ie, if player $1$'s value for this bit is $1$); this continues until the output is determined.  In detail, player $1$ announces the most significant (first) bit of $x_1$.  After player $1$ announces his $j^\mathrm{th}$ bit, if this bit is $0$ and $j<k$, then player $1$ announces his $(j+1)^\mathrm{st}$ bit.  If this bit is $0$ and $j=k$, then the protocol terminates (with, if computing \disjoint, output $1$).  If this bit is $1$, then player $2$ announces the value of his $j^\mathrm{th}$ bit.  If player $2$'s $j^\mathrm{th}$ bit is also $1$, then for \disjoint\ the protocol terminates with output $0$, and for \intersection\ the protocol continues (with $k+1-j$ in the output set); if player $2$'s bit is $0$ and $j<k$, then player $1$ announces his $(j+1)^\mathrm{st}$ bit, while if $j=k$, then the protocol terminates.

\paragraph{Alternating protocol} In the alternating protocol, the role of being the first player to announce the value of a particular bit alternates between the players whenever the first player to announce the value of his $j^\mathrm{th}$ bit announces ``$0$'' (in which case the other player does not announce the value of his corresponding bit).  This continues until the output is determined.  In detail, player $1$ starts by announcing the most significant (first) bit of $x_1$.  After player $i$ announces the value of his $j^\mathrm{th}$ bit, if this bit is $0$ and $j<k$, then the other player announces his $j+1^\mathrm{st}$ bit; if $i$'s $j^\mathrm{th}$ bit is $0$ and $j=k$, the protocol terminates (with output $1$ if computing \disjoint).

If $i$'s $j^\mathrm{th}$ bit is $1$ and the other player had previously announced his $j^\mathrm{th}$ bit (which would necessarily be $1$, else player $i$ would not be announcing his $j^\mathrm{th}$ bit), then the protocol terminates with output $0$ if computing \disjoint, or it continues with the other player announcing his $(j+1)^\mathrm{st}$ bit (and with $k+1-j$ being part of the output set).  If $i$'s $j^\mathrm{th}$ bit is $1$ and the other player had not previously announced his $j^\mathrm{th}$ bit, then the other player announces his $j^\mathrm{th}$ bit; if that bit is $0$, then player $i$ proceeds as above.  If that bit is $1$ and \disjoint\ is being computed, the protocol terminates with output $0$; if the bit is $1$ and \intersection\ is being computed, then player $i$ proceeds as above (and $k+1-j$ will be in the output set).

\subsection{Results}

Table~\ref{tab:results} summarizes our PAR results for the \disjoint\ and \intersection\ problems.  The rows labeled with ``All'' describe bounds for all protocols for that problem (as reflected by the inequalities).  Asymptotic results are for $k\rightarrow\infty$; entries of ``---'' for bounds on subjective PARs indicate that we do not have results beyond those implied by the PARs for specific protocols.
For \intersection, the results for the trivial and $1$-first protocols are shown together; as shown in Lemma~\ref{lem:int-triv1first}, these protocols induce the same tiling, so the PAR results are the same.  All of these results are for average-case objective PARs with respect to the uniform distribution.  These include objective and subjective PARs and the ratio of the subjective PARs.

\begin{table}[htp]
\begin{center}
\small
\begin{tabular}{|c|c|c|c|c|}
  \hline
  Problem & Protocol & Objective PAR & Subjective PAR & Ratio of \\
          &          &               &                & Subj.\ PARs \\ \hline\hline
  \disjoint & All & $\geq \left(\frac{3}{2}\right)^k$ & --- & --- \\ \hline
    & Trivial & $\sim 2^k$ & $\sim 2^k$ & $\sim 2^k$ \\ \hline
    & $1$ First & $\sim2^k$ & $\sim\left(\frac{3}{2}\right)^k$ & $\sim\frac{2}{k}\left(\frac{3}{2}\right)^k$ \\ \hline
    & Alternating & $\sim 2^k$ & $\sim\frac{3+2\sqrt{2}}{2}\left(\frac{1+\sqrt{2}}{2}\right)^k$ & $\sim\sqrt{2}$ \\ \hline \hline
  \intersection & All & $\geq\left(\frac{7}{4}\right)^k$ & --- & --- \\ \hline
    & Trivial/$1$ First &  $\left(\frac{7}{4}\right)^k$  & $\left(\frac{3}{2}\right)^k$ & $\left(\frac{3}{2}\right)^k$ \\ \hline
    & Alternating & $\left(\frac{7}{4}\right)^k$ & $\frac{6}{5}\left(\frac{5}{4}\right)^k$ & $\frac{3}{2}$ \\ \hline
\end{tabular}
\end{center}
\caption{Summary of results.  Asymptotic results are for $k\rightarrow\infty$.}\label{tab:results}
\end{table}

\subsubsection{Discussion of results for \disjoint}

All three protocols have the lowest possible average-case objective PAR for \disjoint.  They also have average-case subjective PARs that are exponential in $k$, although the bases differ.  When considering these protocols (and the tilings they induce as depicted in Sec.~\ref{sec:disjoint}), however, our intuition is that players are much less likely to participate in the trivial and $1$-first protocols (if they do so as player $1$) than they are to participate in the alternating protocol.  This is captured by the comparison of the average-case subjective PAR with respect to the two players in each protocol: In the trivial and $1$-first protocols, the subjective PAR with respect to player $2$ is exponentially worse than the subjective PAR with respect to player $1$; by contrast, in the alternating protocol the subjective PARs differ (asymptotically) by a constant factor.  We do not have any absolute lower bound for the average-case subjective PAR for \disjoint.  However, we conjecture that this grows exponentially.
\begin{conj}
The average-case subjective PAR for \disjoint\ with respect to the uniform distribution grows exponentially in $k$.
\end{conj}

\subsection{Discussion of results for \intersection}

From a high-level perspective, the PAR results for \intersection\ are very similar to those for \disjoint.  As for their \disjoint\ variants, all three protocols have exponentially large average-case objective PAR for \intersection; we show that the average-case objective PAR for \intersection\ is also exponential in $k$, and we conjecture that this bound can be tightened to match the $2^k$ asymptotic growth of the average-case objective PAR for all three of these protocols.

\begin{conj}
The average-case objective PAR for \intersection\ is asymptotic to $2^k$.
\end{conj}

All three protocols also have average-case subjective PARs that are exponential in $k$, although the bases differ.  Our intuition that the alternating protocol is significantly better is not captured by the average-case objective and subjective PARs, but we again see it when we consider the ratio of the subjective PARs: In the trivial and $1$-first protocols, the subjective PAR for player $1$ is exponentially worse than the subjective PAR for player $2$; by contrast, in the alternating protocol the subjective PARs differ by a constant factor of $\frac{3}{2}$.  We do not have any absolute lower bound for the average-case subjective PAR for \intersection.  However, as for \disjoint, we conjecture that this grows exponentially.
\begin{conj}
The average-case subjective PAR for \intersection\ with respect to the uniform distribution grows exponentially in $k$.
\end{conj}

\section{PARs for \disjoint}\label{sec:disjoint}

\subsection{Structure of Protocol-Induced Tilings}

The tiling induced by the trivial protocol is straightforward.  For every input $S_1\neq 0^k$ held by player $1$, there are two monochromatic rectangles in the corresponding row of the input space: $\{(S_1,S_2)|S_2\cap S_1 \neq \emptyset\}$ and $\{(S_1,S_2)|S_2\cap S_1 = \emptyset\}$.  The row corresponding to $S_1=0^k$ forms a single monochromatic rectangle.

\begin{figure}[htp]
\begin{center}
\includegraphics[height=2.5in]{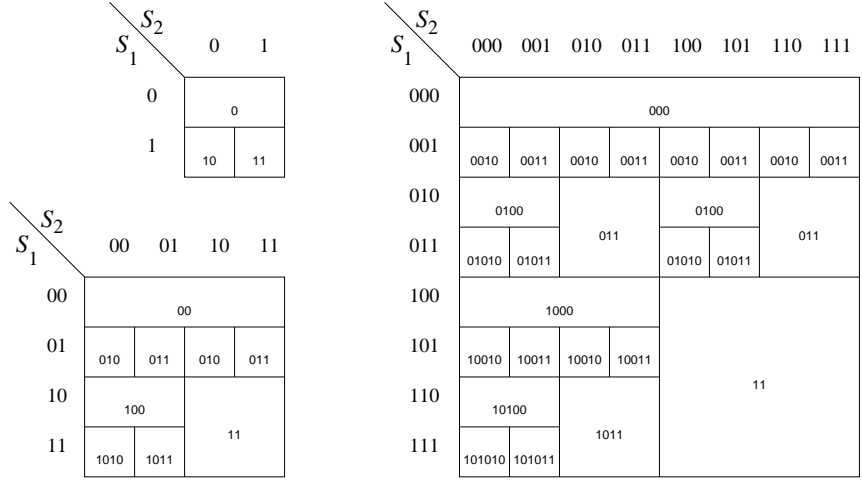}
\caption{Partition of the value space for $k=1$ (top left), $2$ (bottom left), and $3$ (right) induced by the $1$-first protocol for \disjoint; each rectangle is labeled with the transcript output by the protocol when run on inputs in the rectangle.}\label{fig:dis-1f-k1k2k3}
\end{center}
\end{figure}

Figure~\ref{fig:dis-1f-k1k2k3} depicts the $1$-first-protocol-induced tiling of the $1$-, $2$-, and $3$-bit input spaces.  Each tile is labeled with the transcript produced by the protocol on inputs from that tile; note that some tiles are depicted as non-contiguous regions.  When the input space is depicted as in Fig.~\ref{fig:dis-1f-k1k2k3} (\ie, with the possible values of $S_1$ and $S_2$ arranged in increasing lexicographic order from the top-left corner), the tiling of the $k+1$-bit input space induced by the $1$-first protocol can be obtained as follows.  Let $T_k$ be the $1$-first-protocol-induced tiling of the $k$-bit input space.  The top-left and top-right quadrants of $T_{k+1}$ are copies of $T_k$; in each of these quadrants, a trace in $T_{k+1}$ is the corresponding trace in $T_k$ prepended with $0$.  The bottom-left quadrant of $T_{k+1}$ is another copy of $T_k$, with each trace in this part of $T_{k+1}$ being obtained by prepending $10$ to the corresponding trace in $T_k$.  The bottom-right quadrant of $T_{k+1}$ is a single rectangle whose trace is $11$.

\begin{figure}[htp]
\begin{center}
\includegraphics[height=2.5in]{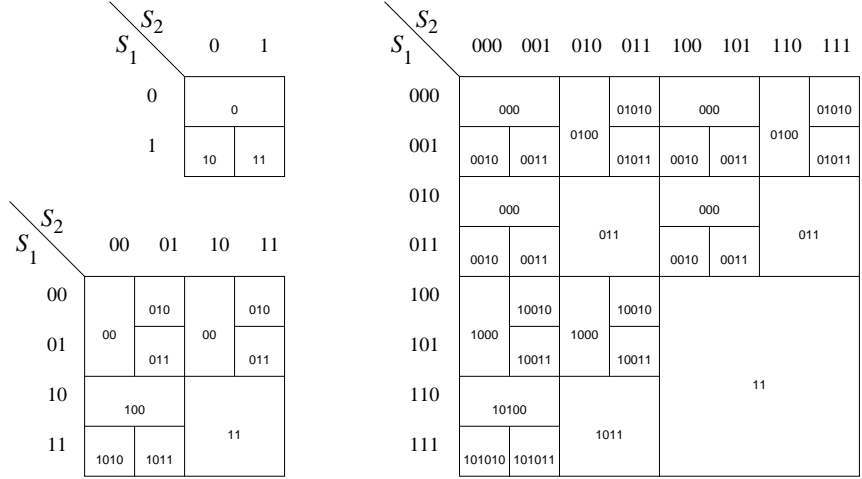}
\caption{Partition of the value space for $k=1$ (top left), $2$ (bottom left), and $3$ (right) induced by the alternating protocol for \disjoint; each rectangle is labeled with the transcript output by the protocol when run on inputs in the rectangle.}\label{fig:dis-alt-k1k2k3}
\end{center}
\end{figure}

Figure~\ref{fig:dis-alt-k1k2k3} shows the partition of the $1$-, $2$-, and $3$-bit input spaces induced by the alternating protocol; each induced rectangle is labeled with the corresponding transcript (note that some rectangles appear as non-contiguous regions in the figure).  If we denote by $T_k$ the tiling of the $k$-bit space induced by the alternating protocol as depicted in Fig.~\ref{fig:dis-alt-k1k2k3}, then the bottom-left quadrant of $T_{k+1}$ has the same structure as $T_k$, with the transcript for a tile in $T_{k+1}$ obtained by prepending $10$ to the transcript for the corresponding tile in $T_k$.  Each of the top quadrants has the same structure as the reflection of $T_k$ across the top-left-to-bottom-right diagonal; the corresponding rectangles in these quadrants actually form single rectangles, and the associated transcript is obtained by prepending $0$ to the transcript for the corresponding rectangle in $T_k$.  Finally, the bottom-right quadrant is a single rectangle that always has the transcript $11$.

\subsection{Objective PAR}

\subsubsection{Objective PAR for the \disjoint\ problem}

\begin{lemma}
In the ideal partition induced by \disjoint, at least $2^k$ rectangles are required to tile the region $f^{-1}(1)$.
\end{lemma}
\begin{proof}
As shown in, \eg,~\cite{KN97}, the $2^k$ input pairs $(S,\{1,\ldots,k\}\setminus S)$ form a ``fooling set''---no two of these input pairs can belong to the same monochromatic rectangle.
\end{proof}

\begin{cor}\label{cor:dis-pardis}
The average-case objective PAR of \disjoint\ with respect to the uniform distribution is at least $\left(\frac{3}{2}\right)^k$.
\end{cor}
\begin{proof}
The contribution to the sum in Eq.~\ref{eq:pargen} from the protocol-induced tiles $S\subset f^{-1}(1)$ must be at least $2^k\cdot 3^k$, so the average-case objective PAR with respect to the uniform distribution is at least $\left(\frac{3}{2}\right)^k$.
\end{proof}

\subsubsection{Objective PAR for specific protocols}

\begin{lemma}\label{lem:dis-par}
If a protocol $P$ for \disjoint\ tiles $f^{-1}(1)$ with $2^k$ tiles and tiles $f^{-1}(0)$ with $2^k - 1$ tiles, then the average-case objective PAR of $P$ with respect to the uniform distribution equals
\[
2^k -1 + \left(\frac{3}{4}\right)^k.
\]
\end{lemma}
\begin{proof}
By the argument for Cor.~\ref{cor:dis-pardis}, the contribution to this PAR value from those $S\subset f^{-1}(1)$ is $\left(\frac{3}{2}\right)^k$.  The contribution to this PAR value from those $S\subset f^{-1}(0)$ is $4^{-k}\cdot (4^k-3^k)\cdot (2^k-1)$.  Summing these together, we obtain the claimed value.
\end{proof}

\begin{prop}
The average-case objective PAR of the trivial protocol for \disjoint\ with respect to the uniform distribution is
\[
2^k -1 + \left(\frac{3}{4}\right)^k.
\]
\end{prop}
\begin{proof}
The trivial protocol tiles $f^{-1}(1)$ with $2^k$ tiles (one for each set $S_1$ that player $1$ might have), and it tiles $f^{-1}(0)$ with $2^k - 1$ tiles (one for each non-empty set $S_1$ that player $1$ might have).  We may then apply Lemma~\ref{lem:dis-par}.
\end{proof}

\begin{prop}
The average-case objective PAR of the $1$-first protocol for \disjoint\ with respect to the uniform distribution is
\[
2^k -1 + \left(\frac{3}{4}\right)^k.
\]
\end{prop}
\begin{proof}
The protocol-induced tiles of $f^{-1}(1)$ correspond bijectively to the $2^k$ possible protocol transcripts $\{0,10\}^k$, while the protocol-induced tiles of $f^{-1}(0)$ correspond bijectively to the $2^k - 1$ possible protocol transcripts $\left\{\{0,10\}^i\times \{11\}\right\}_{i=0}^{k-1}$.  We may then apply Lemma~\ref{lem:dis-par}.
\end{proof}

\begin{prop}
The average-case objective PAR of the alternating protocol for \disjoint\ with respect to the uniform distribution is
\[
2^k -1 + \left(\frac{3}{4}\right)^k.
\]
\end{prop}
\begin{proof}
The protocol-induced tiles of $f^{-1}(1)$ correspond bijectively to the $2^k$ possible protocol transcripts $\{0,10\}^k$, while the protocol-induced tiles of $f^{-1}(0)$ correspond bijectively to the $2^k - 1$ possible protocol transcripts $\left\{\{0,10\}^i\times \{11\}\right\}_{i=0}^{k-1}$.  We may then apply Lemma~\ref{lem:dis-par}.
\end{proof}

\subsection{Subjective PAR}

\subsubsection{Subjective PAR for the trivial protocol}

\begin{prop}
The average-case PAR with respect to player $1$ of the trivial protocol for \disjoint\ is $1$.  The average-case PAR with respect to player $2$ of the trivial protocol for \disjoint, and thus the average-case subjective PAR for the protocol, is
\[
2^k -2\left(\frac{3}{2}\right)^k + 2\left(\frac{5}{4}\right)^k \sim 2^k\qquad (k\rightarrow\infty).
\]
\end{prop}
\begin{proof}
The $1$-partition induced by the trivial protocol is exactly the ideal $1$-partition, from which the first claim follows.

The $2$-partition induced by the trivial protocol distinguishes between every pair of distinct inputs.  To compute the average-case PAR with respect to player $2$, we use $\svn_k$ and $\svy_k$ to denote the contributions (in the $k$-bit version of the problem) to the sum in Eq.~\ref{eq:pargen} from tiles in $f^{-1}(0)$ and $f^{-1}(1)$, respectively, so the average-case PAR with respect to player $2$ is then $\left(\svn_k+\svy_k\right)/4^k$.

Let $S$ be a $2$-rectangle induced by the trivial protocol in the $k+1$-bit value space (so $S$ is $1\times 1$).  If $S$ is in either the bottom-left or the top-left quadrant, then the size of the ideal rectangle containing $S$ is twice the size of the ideal rectangle that contains the corresponding induced rectangle in the $k$-bit value space (\ie, the point in the $k$-bit space obtained by omitting the first bit of each input in $S$ when the value space is depicted as in Fig.~\ref{fig:disk3}).  This holds regardless of whether $S\subset f^{-1}(0)$ or $S\subset f^{-1}(1)$.  If $S$ is in the top-right quadrant and $S\subset f^{-1}(1)$, then the size of the ideal rectangle containing $S$ is the same as that of the ideal rectangle containing the corresponding input in the $k$-bit value space; note that the bottom-right quadrant does not contain any points in $f^{-1}(1)$.  If $S$ is in the top-right quadrant and $S\subset f^{-1}(0)$, then the size of the ideal rectangle containing $S$ is that of the ideal rectangle containing the corresponding input in the $k$-bit value space plus $2^k$; the extra contribution of $2^k$ is added on for each of the $4^k-3^k$ protocol-induced $2$-rectangles in the top-right quadrant.  If $S$ is in the bottom-right quadrant (so that it is necessarily contained in $f^{-1}(0)$), then the size of the ideal rectangle containing $S$ is at least $2^k$ (the part of the containing rectangle that is in the bottom-right quadrant); the amount by which this exceeds $2^k$ equals the size of the ideal $2$-rectangle (for $f^{-1}(0)$) containing the corresponding point in the $k$-bit value space.  In particular, each of the $2$-rectangles for the $k$-bit value space is counted for exactly $2^k$ induced rectangles in the bottom-right quadrant, so the entire excess contribution is $2^k (4^k-3^k)$.

We thus obtain the following recurrences (the terms are grouped by quadrant, clockwise from the bottom left).
\begin{eqnarray*}
\svn_{k+1} &=& 2\svn_k + 2\svn_k + \left(\svn_k + 2^k(4^k-3^k)\right) + \left(4^k\cdot 2^k + 2^k\cdot(4^k - 3^k)\right)\quad \svn_1 = 1\\
\svy_{k+1} &=& 2\svy_k + 2\svy_k + \svy_k + 0\qquad\qquad \svy_1 = 5
\end{eqnarray*}
From these, we obtain $\svy_k = 5^k$ and
\[
\svn_{k} = 2^{3k} - 2^{1 + k}3^k + 5^k,
\]
from which it follows that the average-case subjective PAR with respect to player $2$ (and thus for the trivial protocol) is
\[
\frac{1}{4^k}(8^k -(2^{k+1}3^k) + 2\cdot 5^k) = 2^k -2\left(\frac{3}{2}\right)^k + 2\left(\frac{5}{4}\right)^k.
\]
\end{proof}

\begin{cor}
If $\PAR^\mathrm{trivial}_i$ denotes the average-case PAR w.r.t. $i$ of the trivial protocol for \disjoint\ w.r.t. the uniform distribution, then
\[
\frac{\PAR^\mathrm{trivial}_2}{\PAR^\mathrm{trivial}_1} \sim 2^k\qquad (k\rightarrow\infty).
\]
\end{cor}

\subsubsection{Subjective PAR for the $1$-first protocol}

\begin{theorem}
The average-case PAR with respect to player $1$ of the $1$-first protocol for \disjoint\ with respect to the uniform distribution is
\[
    \frac{k}{2} - \frac{k}{3}\left(\frac{3}{4}\right)^k + \left(\frac{3}{4}\right)^k \sim \frac{k}{2}\qquad (k\rightarrow\infty).
\]
The average-case PAR with respect to player $2$ of the $1$-first protocol for \disjoint\ with respect to the uniform distribution is
\[
    \left(\frac{3}{2}\right)^k + \frac{1}{2}\left(\frac{5}{4}\right)^k - 1 + \frac{1}{2}\left(\frac{3}{4}\right)^k \sim \left(\frac{3}{2}\right)^k\qquad (k\rightarrow\infty).
\]
\end{theorem}
\begin{proof}
To compute the average-case PAR with respect to player $1$, we use $\shn_k$ and $\shy_k$ to denote the contributions (in the $k$-bit version of the problem) to the sum in Eq.~\ref{eq:pargen} from the $1$-induced tiles in $f^{-1}(0)$ and $f^{-1}(1)$, respectively, so the average-case PAR with respect to player $1$ is then $\left(\shn_k+\shy_k\right)/4^k$.

Let $S\subset f^{-1}(1)$ be a $1$-rectangle induced by the $1$-first protocol in the $(k+1)$-bit value space.  If $S$ is in the bottom-left quadrant, the ideal $1$-rectangle containing $S$ is the same size as the ideal rectangle that contains $S$ in the $k$-bit value space (because there are no inputs in the bottom-right quadrant in $f^{-1}(1)$).  If $S$ is in one of the top quadrants, then the ideal $1$-rectangle containing $S$ is twice the size of the rectangle containing the rectangle that corresponds to $S$ in the partition of the $k$-bit value space.  Observe that each point in the top-left quadrant is in the same rectangle as the corresponding point in the top-right quadrant; in particular, this means that the induced $1$-rectangles in the top two quadrants correspond bijectively to the induced $1$-rectangles in the $k$-bit value space.  $S$ cannot be in the bottom-right quadrant, which contains no points in $f^{-1}(1)$.  We thus have (separating the contributions of the bottom-left, top, and bottom-right quadrants)
\[
\shy_{k+1} = \shy_k + 2\shy_k + 0 = 3\shy_k.
\]
By inspection, $\shy_1 = 1+2+0 = 3$; so $\shy_k = 3^k$.

Now let $S\subset f^{-1}(0)$ be a $1$-rectangle induced by the $1$-first protocol in the $(k+1)$-bit value space.  If $S$ is in the bottom-left quadrant, then the size of the ideal $1$-rectangle containing $S$ equals the size of the ideal $1$-rectangle containing $S$ in the $k$-bit value space plus $2^k$ (because all of the inputs in the bottom-right quadrant in the same $1$-rectangle as $S$ are in the same ideal $1$-rectangle as $S$).  If $\nh_k$ denotes the number of induced $1$-rectangles $S\subset f^{-1}(0)$ in the bottom-left quadrant (this is the same as the total number of such $1$-rectangles in the $k$-bit space), then the total extra contribution is $2^k\nh_k$.  If $S$ is in the top two quadrants, the same arguments as before apply.  If $S$ is in the bottom-right quadrant (so that the size of $S$ is $2^k$), then the ideal $1$-rectangle containing $S$ has size $2^k$ plus the size of whatever part of the ideal $1$-rectangle lies in the bottom-left quadrant.  If we sum over all $2^k$ rectangles $S$ in the bottom-right quadrant, the extra contribution from the bottom-left quadrant equals the total size of $f^{-1}(0)$ in the $k$-bit value space, \ie, $4^k-3^k$.  This leads to (again separating the contributions of the bottom-left, top, and bottom-right quadrants)
\[
\shn_{k+1} = (\shn_k + 2^k\nh_k) + 2\shn_k + ((4^k-3^k)+4^k).
\]
By inspection, $\shn_1 = 1$.  Because the bottom-left quadrant is a copy of the tiling of the $k$-bit space, the top two quadrants have the same number of rectangles, and the bottom-right quadrant has $2^k$ $1$-rectangles, we have
\[
\nh_{k+1} = \nh_k + \nh_k + 2^k,
\]
with $\nh_1 = 1$.  From this, we obtain
\[
\nh_k = k\cdot 2^{k-1},
\]
which we then use to obtain
\[
\shn_k = \frac{k}{6}\left(3\cdot 4^k - 2\cdot 3^k\right).
\]
Using $\PAR_1$ to denote the average-case PAR with respect to $1$, we have
\begin{eqnarray*}
\PAR_1 &=& \frac{1}{4^k}(\shn_k + \shy_k)\\
    &=& \frac{1}{4^k}\left(\frac{k}{6}3\cdot 4^k - \frac{k}{6}2\cdot 3^k + 3^k\right)\\
    &=& \frac{k}{2} - \frac{k}{3}\left(\frac{3}{4}\right)^k + \left(\frac{3}{4}\right)^k
\end{eqnarray*}
as claimed.

We now turn to the computation of the average-case PAR with respect to player $2$.  We use $\svn_k$ and $\svy_k$ to denote the contributions (in the $k$-bit version of the problem) to the sum in Eq.~\ref{eq:pargen} from the $2$-induced tiles in $f^{-1}(0)$ and $f^{-1}(1)$, respectively, so the average-case PAR with respect to player $2$ is then $\left(\svn_k+\svy_k\right)/4^k$.

Let $S\subset f^{-1}(1)$ be a $2$-rectangle induced by the $1$-first protocol in the $(k+1)$-bit value space.  If $S$ is in the bottom-left quadrant, the ideal $2$-rectangle containing $S$ is twice as big as the ideal $2$-rectangle that contains $S$ in the $k$-bit value space.  The same holds true if $S$ is in the top-left quadrant.  If $S$ is in the top-right quadrant, the ideal $2$-rectangle containing $S$ is the same size as in the $k$-bit value space.  Finally, the bottom-right quadrant does not contain any values in $f^{-1}(1)$.  Thus, we have (again listing contributions clockwise from the bottom-left quadrant)
\[
\svy_{k+1} = 2\svy_k + 2\svy_k + \svy_k + 0 = 5\svy_k.
\]
By inspection, $\svy_1 = 5$; so, $\svy_k = 5^k$.

Now let $S\subset f^{-1}(0)$ be a $2$-rectangle induced by the $1$-first protocol in the $(k+1)$-bit value space.  If $S$ is in the bottom-left or top-left quadrant, the ideal $2$-rectangle containing $S$ is twice as big as in the $k$-bit value space.  If $S$ is in the top-right quadrant, the size of the ideal $2$-rectangle containing $S$ equals $2^k$ plus the size of the ideal $2$-rectangle that contains $S$ in the $k$-bit value space.  Finally, if we sum over all $S$ in the bottom-right quadrant, the total sizes of the ideal $2$-rectangles containing these $S$ is $2^k\cdot 2^k$ plus the total size of $f^{-1}(0)$ in the $k$-bit value space.  Combining all of these relations, and using $\nv_k$ to denote the number of $2$-rectangles in $f^{-1}(0)$ in the $k$-bit value space, we have
\[
\svn_{k+1} = 2\svn_k + 2\svn_k + (\svn_k + 2^k\cdot \nv_k) + (4^k + 4^k-3^k).
\]
(As above, contributions are grouped by quadrant clockwise from the bottom right.)\ \ We also have
\[
\nv_{k+1} = \nv_k + \nv_k + \nv_k + 2^k = 3\nv_k + 2^k.
\]
By inspection, $\svn_1 = 1$ and $\nv_1=1$.  From this, we obtain
\[
\nv_k = 3^k - 2^k
\]
and then
\[
\svy_k = -4^k + \frac{1}{2}3^k + 6^k - \frac{1}{2}5^k.
\]
Using $\PAR_2$ to denote the average-case PAR with respect to $2$, we have
\begin{eqnarray*}
\PAR_2 &=& \frac{1}{4^k} (\svn_k + \svy_k)\\
    &=& \frac{1}{4^k} \left(5^k + 6^k - \frac{1}{2}5^k - 4^k + \frac{1}{2}3^k \right)\\
    &=& \left(\frac{3}{2}\right)^k + \frac{1}{2}\left(\frac{5}{4}\right)^k - 1 + \frac{1}{2}\left(\frac{3}{4}\right)^k
\end{eqnarray*}
as claimed.
\end{proof}

\begin{cor}
The average-case subjective PAR of the $1$-first protocol for \disjoint\ with respect to the uniform distribution is
\[
    \left(\frac{3}{2}\right)^k + \frac{1}{2}\left(\frac{5}{4}\right)^k - 1 + \frac{1}{2}\left(\frac{3}{4}\right)^k \sim \left(\frac{3}{2}\right)^k\qquad (k\rightarrow\infty).
\]
\end{cor}

\begin{cor}
If $\PAR^\mathrm{1-first}_i$ denotes the average-case PAR w.r.t. $i$ of the $1$-first protocol for \disjoint\ w.r.t. the uniform distribution, then
\[
\frac{\PAR^\mathrm{1-first}_2}{\PAR^\mathrm{1-first}_1} \sim \frac{2}{k}\left(\frac{3}{2}\right)^k\qquad (k\rightarrow\infty).
\]
\end{cor}

\subsubsection{Subjective PAR for the alternating protocol}

We let $\paratio^i_k$ denote the PAR w.r.t.\ $i$ for the alternating protocol for \disjoint.  We let $\hy_k$ and $\vy_k$ be the contributions of $f^{-1}(1)$ to the sums analogous to that in Eq.~\ref{eq:pargen} for objective PAR, \ie,
\[
\hy_k = \sum_{S\subseteq f^{-1}(1)} |R^I(S)| \qquad \vy_k = \sum_{T\subseteq f^{-1}(1)} |R^I(T)|,
\]
where the sum for $\hy_k$ is taken over protocol-induced ``horizontal'' rectangles $S$ (in the induced $1$-partition) on which $f$ takes the value $1$, and the sum for $\vy_k$ is taken over protocol-induced ``vertical'' rectangles $T$ (in the induced $2$-partition) on which $f$ takes the value $1$.  Using the structure of the induced tiling,
we may obtain recurrences for $\hy_k$ and $\vy_k$ as follows.
\begin{eqnarray}
\hy_k &=& \hy_{k-1} + 2\vy_{k-1} + 0\qquad\qquad \hy_1 = 3\label{eq:hyk}\\
\vy_k &=& 2\vy_{k-1} + \left(2\hy_{k-1}+\hy_{k-1}\right) + 0\qquad\qquad \vy_1 = 5\label{eq:vyk}
\end{eqnarray}
In each recurrence, the first summand is the contribution from the bottom-left quadrant, the second summand is the contribution from the two top quadrants, and the third summand is the contribution from the bottom-right quadrant.  From these recurrences, we obtain $\hy_k = \frac{4}{5}2^{2k} + \frac{(-1)^k}{5}$ and $\vy_k = \frac{6}{5}2^{2k} - \frac{(-1)^k}{5}$.

We define $\hn_k$ and $\vn_k$ analogously to capture the contributions of $f^{-1}(0)$ to the sums under consideration; we will also keep track of the number of tiles in the $1$- and $2$-induced partitions on which $f$ takes the value $0$ (the ``horizontal'' and ``vertical'' tiles, which we denote as $\nh_k$ and $\nv_k$, respectively).

We start with the following recurrences for $\nh_k$ and $\nv_k$.
\begin{eqnarray*}
\nh_{k+1} &=& \nh_{k} + \nv_{k} + 2^{k} \qquad\qquad \nh_1 = 1\\
\nv_{k+1} &=& \nv_{k} + \left(\nh_{k} + \nh_{k}\right) + 2^{k} \qquad\qquad \nv_1 = 1
\end{eqnarray*}
From these, we obtain
\begin{eqnarray*}
\nh_k &=& -2^{k+1} + (1 - 3/(2\sqrt{2}))\cdot (1 -\sqrt{2})^k + ((1 + \sqrt{2})^k\cdot (4 + 3\sqrt{2}))/4\\
\nv_k &=& -3\cdot 2^k + (1 - \sqrt{2})^k\cdot (3/2 - \sqrt{2}) + (1 + \sqrt{2})^k \cdot (3/2 + \sqrt{2})
\end{eqnarray*}

We obtain the following recurrences for $\hn_k$ and $\vn_k$.
\begin{eqnarray*}
\hn_{k+1} &=& \left(\hn_k + \nh_k\cdot 2^k\right)+2\vn_k + \left(4^k+4^k-3^k\right)\qquad\qquad \hn_1 = 1\\
\vn_{k+1} &=& 2\vn_k + 2\hn_k + \left(\hn_k + \nh_k\cdot 2^k \right)+\left(4^k + 4^k - 3^k\right) \qquad\qquad \vn_1 = 1
\end{eqnarray*}
From these, we may obtain
\begin{multline*}
\hn_k = \frac{1}{20\sqrt{2}}\bigg(5\cdot 2^{k+1}\cdot (1 - \sqrt{2})^k\cdot (-3 + 2\sqrt{2}) + 5\cdot 2^{k+1}\cdot (1 + \sqrt{2})^k\cdot (3 + 2\sqrt{2})\\ + \sqrt{2}((-1)^k - 7\cdot 2^{2k+3} + 5\cdot 3^{k+1})\bigg)
\end{multline*}
\begin{multline*}
\vn_k = \frac{1}{20}\bigg(-(-1)^k + 25\cdot 3^k - 21\cdot 4^{k+1} - 5\cdot 2^{k+1}(1 - \sqrt{2})^k(-3 + 2\sqrt{2})\\ + 5\cdot 2^{k+1}(1 + \sqrt{2})^k(3 + 2\sqrt{2})\bigg)
\end{multline*}

We may now compute the PAR with respect to each of the two players as
\[
\paratio^\mathrm{alt}_1(k) = \frac{\hn_k + \hy_k}{2^{2k}}\qquad\mathrm{and}\qquad\paratio^\mathrm{alt}_2(k) = \frac{\vn_k + \vy_k}{2^{2k}}.
\]

\begin{theorem}
The average-case PAR with respect to player $1$ of the alternating protocol for \disjoint\ with respect to the uniform distribution is
\begin{multline*}
\paratio^\mathrm{alt}_1(k) = \frac{1}{4^{k+1}}\bigg((-1)^k - 2^{2k+3} + 3^{k+1}\\ + (4-3\sqrt{2})(2-2\sqrt{2})^k + (2+2\sqrt{2})^k(4+3\sqrt{2})\bigg)\\
   \sim \frac{4+3\sqrt{2}}{4}\left(\frac{1+\sqrt{2}}{2}\right)^k \qquad (k\rightarrow\infty)
\end{multline*}
The average-case PAR with respect to player $2$ of the alternating protocol for \disjoint\ with respect to the uniform distribution is
\begin{multline*}
\paratio^\mathrm{alt}_2(k) = \frac{1}{4^{k+1}}\bigg(-(-1)^k + 5\cdot 3^k - 3\cdot 4^{k+1}\\ + 2^{k+1}(3-2\sqrt{2})(1-\sqrt{2})^k + 2^{k+1}(3+2\sqrt{2})(1+\sqrt{2})^k\bigg)\\
    \sim \frac{3+2\sqrt{2}}{2}\left(\frac{1+\sqrt{2}}{2}\right)^k \qquad (k\rightarrow \infty)
\end{multline*}
\end{theorem}

\begin{cor}
The average-case subjective PAR of the alternating protocol for \disjoint\ with respect to the uniform distribution is
\begin{multline*}
\frac{1}{4^{k+1}}\bigg(-(-1)^k + 5\cdot 3^k - 3\cdot 4^{k+1}
    + 2^{k+1}(3-2\sqrt{2})(1-\sqrt{2})^k + 2^{k+1}(3+2\sqrt{2})(1+\sqrt{2})^k\bigg)\\
    \sim \frac{3+2\sqrt{2}}{2}\left(\frac{1+\sqrt{2}}{2}\right)^k \qquad (k\rightarrow \infty)
\end{multline*}
\end{cor}

\begin{cor}
If $\PAR^\mathrm{alt}_i(k)$ denotes the average-case PAR w.r.t. $i$ of the $1$-first protocol for \disjoint\ w.r.t. the uniform distribution, then
\[
\frac{\PAR^\mathrm{alt}_2(k)}{\PAR^\mathrm{alt}_1(k)} \sim \sqrt{2}\qquad (k\rightarrow\infty).
\]
\end{cor}

\section{PARs for \intersection}\label{sec:intersection}

\subsection{Structure of Protocol-Induced Tilings}

First, we observe that for \intersection, the trivial and $1$-first protocols induce the same tiling.
\begin{lemma}\label{lem:int-triv1first}
The tilings induced by the trivial and $1$-first protocols for \intersection\ are identical.
\end{lemma}
\begin{proof}
Given two input pairs $(S_1,S_2)$ and $(T_1,T_2)$, each of these protocols cannot distinguish between the pairs if and only if (1) $S_1=T_1$ and (2) $S_2$ and $T_2$ differ only on elements that are not in $S_1=T_1$.
\end{proof}

\begin{figure}[htp]
\begin{center}
\includegraphics[height=2.5in]{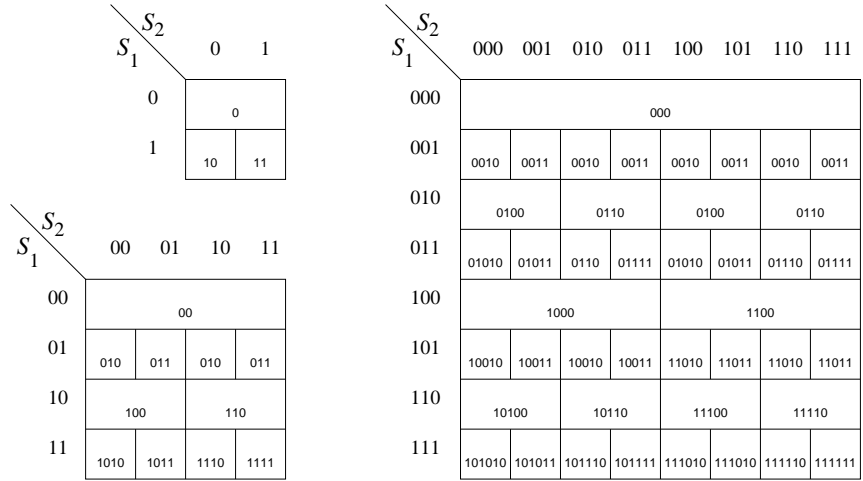}
\caption{Partition of the value space for $k=1$ (top left), $2$ (bottom left), and $3$ (right) induced by the trivial and $1$-first protocols for \intersection; each rectangle is labeled with the transcript output by the protocol when run on inputs in the rectangle.}\label{fig:int-1f-k1k2k3}
\end{center}
\end{figure}

Figure~\ref{fig:int-1f-k1k2k3} depicts the tilings of the $1$-, $2$-, and $3$-bit value spaces induced by the trivial and $1$-first protocols for \intersection.  If we denote by $T_k$ the $1$-first-protocol-induced tiling of the $k$-bit input space, then when we depict $T_{k+1}$ as in Fig.~\ref{fig:int-1f-k1k2k3}, the bottom-left quadrant is $10T_k$ (\ie, the $k$-bit tiling with $10$ prepended to each transcript), each of the top quadrants is $0T_k$, and the bottom-right quadrant is $11T_k$.

\begin{figure}[htp]
\begin{center}
\includegraphics[height=2.5in]{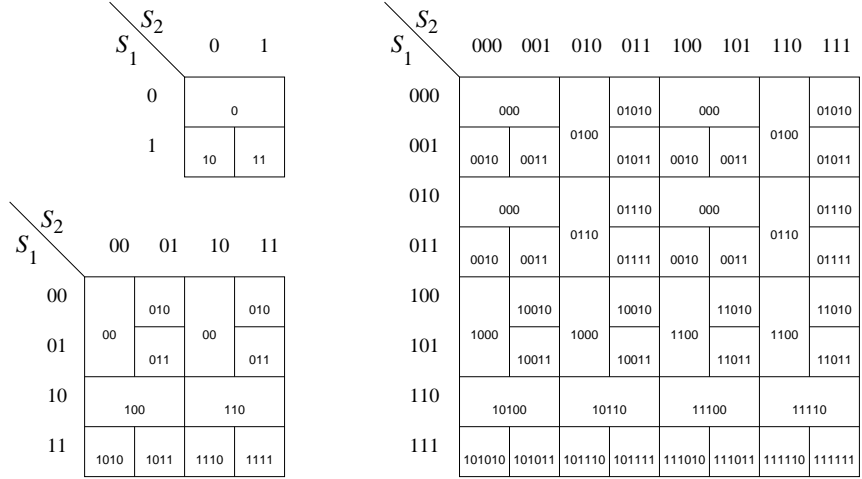}
\caption{Partition of the value space for $k=1$ (top left), $2$ (bottom left), and $3$ (right) induced by the alternating protocol for \intersection; each rectangle is labeled with the transcript output by the protocol when run on inputs in the rectangle.}\label{fig:int-alt-k1k2k3}
\end{center}
\end{figure}

Figure~\ref{fig:int-alt-k1k2k3} depicts the tilings of the $1$-, $2$-, and $3$-bit value spaces induced by the alternating protocol for \intersection.  If we denote by $T_k$ the alternating-protocol-induced tiling of the $k$-bit value space and depict $T_{k+1}$ as in Fig.~\ref{fig:int-alt-k1k2k3}, the bottom-left quadrant is $10T_k$ (\ie, the $k$-bit tiling with $10$ prepended to each transcript), each of the top quadrants is $0T_{k}^\mathsf{T}$ (\ie, the $k$-bit tiling reflected across the top-left--bottom-right diagonal), and the bottom-right quadrant is $11T_k$.

\subsection{Objective PAR}

\subsubsection{Lower bound}

We obtain the following result for the average-case objective PAR of the \intersection\ problem.
\begin{theorem}
The average-case objective PAR of the \intersection\ problem with respect to the uniform distribution is $\left(\frac{7}{4}\right)^k$.
\end{theorem}
\begin{proof}
We show that $\PAR_{k+1} = \frac{7}{4}\PAR_k$ and that $\PAR_1 = \frac{7}{4}$.

Using Eq.~\ref{eq:pargen}, we may write $\PAR_{k+1}$ as
\begin{equation}
\PAR_{k+1} = \frac{1}{2^{2(k+1)}} \left(\sum_{R=f^{-1}(0\ldots)} |R^I(R)| + \sum_{R=f^{-1}(1\ldots)} |R^I(R)|\right),\label{eq:opar:int1}
\end{equation}
where the first sum is over induced rectangles $R$ in which the intersection set does not contain $k+1$ (\ie, the encoding of the set starts with 0) and the second sum is over induced rectangles $R$ in which the intersection set does contain this element.  Observe that the ideal monochromatic partition of the region corresponding to inputs in which $k+1\in S_1\cap S_2$ (the bottom-right quadrant when depicted as in Fig.~\ref{fig:intk3}) has the same structure as the ideal monochromatic partition of the entire space when only $k$ elements are used.  Similarly, the three regions corresponding to $k+1\notin S_1\cup S_2$ (top-left quadrant), $k+1\in S_1\setminus S_2$ (bottom-left quadrant), and $k+1\in S_2\setminus S_1$ (top-right quadrant) all have this same structure, although each input in these regions belongs to the same monochromatic region as the corresponding inputs in the other two quadrants.

The first observation allows us to rewrite Eq.~\ref{eq:opar:int1} as
\begin{equation}
\PAR_{k+1} = \frac{1}{4}\left(\frac{1}{2^{2k}}\sum_{R=f^{-1}(0\ldots)} |R^I(R)|\right) + \frac{1}{4}\PAR_k.\label{eq:opar:int2}
\end{equation}
We now turn to rewriting the term in parentheses.

Consider an input $(0x_1,0x_2)\in f^{-1}(0x)$ (\ie, $x,x_i\in\{0,1\}^k$ and $x_1\cap x_2 = x$) in the top-left quadrant of the $(k+1)$-bit input space (when depicted as in Fig.~\ref{fig:intk3}).  In any monochromatic tiling of this space, $(0x_1,0x_2)$ may be in the same tile as at most one of the inputs $(0x_1,1x_2)$ (top-right quadrant) and $(1x_1,0x_2)$ (bottom-left quadrant)---if both $(0x_1,1x_2)$ and $(1x_1,0x_2)$ were in the same tile, then $(1x_1,1x_2)\in f^{-1}(1x)$ would also be in this tile, violating monochromaticity.  If $a_x$ is the minimum number of monochromatic tiles needed to tile the region $f^{-1}(x)$ in the $k$-bit input space, then at least $2 a_x$ monochromatic tiles are needed to tile the region $f^{-1}(0x)$ in the $(k+1)$-bit input space.  For any $x\in\{0,1\}^k$, the size of the ideal monochromatic region $f^{-1}(0x)$ is $3$ times the size of the monochromatic region $f^{-1}(x)$ in the ideal partition of the input space for $k$-element sets.  Thus the contribution to the sum (for $\PAR_{k+1}$) in Eq.~\ref{eq:opar:int1} of the rectangles $R$ in $f^{-1}(0x)$ is $6$ times the contributions of the contribution to the sum (for $\PAR_k$) of the rectangles $R$ in $f^{-1}(x)$.  This allows us to rewrite Eq.~\ref{eq:opar:int2} as
\[
\PAR_{k+1} = \frac{6}{4}\PAR_k + \frac{1}{4}\PAR_k.
\]
Finally, the ideal partition for the \intersection\ problem with $k=1$, shown in Fig.~\ref{fig:intk1}, requires at least $2$ tiles for the region (of size $3$) corresponding to an empty intersection and a single tile for the region (of size $1$) corresponding to a non-empty intersection.  This immediately gives the initial condition
\[
\PAR_1 = \frac{1}{2^{2}}\left(3+3+1\right) = \frac{7}{4}.
\]
\begin{figure}[htp]
\begin{center}
\includegraphics[height=.5in]{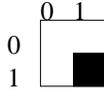}
\end{center}
\caption{Ideal partition for the \intersection\ problem with $k=1$.}\label{fig:intk1}
\end{figure}
\end{proof}

\subsubsection{Objective PAR for the trivial and $1$-first protocols}

\begin{prop}\label{prop:int-ob-triv}
The average-case objective PAR for the trivial and $1$-first protocols for the \intersection\ problem equals $\left(\frac{7}{4}\right)^k$.
\end{prop}
\begin{proof}
Consider the tiling $T_{k+1}$ of the $(k+1)$-bit value space induced by these protocols.  Any tile $S$ in $T_k$ has $3$ corresponding tiles in $T_{k+1}$: the tile whose transcript (in the $1$-first protocol) is $10S$, in the bottom-left quadrant; the tile whose transcript is $0S$, which spans the top two quadrants; and the tile whose transcript is $11S$, which is in the bottom-right quadrant.  The ideal monochromatic region that contains $0S$ and $10S$ (the same region contains both) in the $(k+1)$-bit value space is $3$ times the size of the ideal monochromatic region that contains $S$ in the $k$-bit value space; the ideal monochromatic region that contains $11S$ is the same size as the ideal monochromatic region that contains $S$.  Thus, we have that $\paratio_{k+1} = \frac{7}{4}\paratio_k$.  By inspection, $\paratio_1 = \frac{7}{4}$, finishing the proof.
\end{proof}

\subsubsection{Objective PAR for the alternating protocol}

Although the recursive tiling structure induced by the alternating protocol is slightly different than that induced by the trivial and $1$-first protocols, the argument from the proof of Prop.~\ref{prop:int-ob-triv} applies essentially unchanged.  In particular, even though the structure is different, the tiles in $T_{k+1}$ corresponding to a tile $S$ in $T_k$ are: one tile in the bottom-left quadrant; one tile that spans the top two quadrants; and one tile in the bottom-right quadrant.  Thus, we again have $\paratio_{k+1} = \frac{7}{4}\paratio_k$.  Again, we also have $\paratio_1 = \frac{7}{4}$, giving us the following proposition.
\begin{prop}\label{prop:int-ob-alt}
The average-case objective PAR for the alternating protocol for the \intersection\ problem equals $\left(\frac{7}{4}\right)^k$.\hfil \qed
\end{prop}

\subsection{Subjective PAR}

\subsubsection{Subjective PAR for the trivial and $1$-first protocols}

\begin{remark}
The contribution from $f^{-1}(\emptyset)$ is as for \disjoint.  What about the contribution for $f^{-1}(\neq\emptyset)$?
\end{remark}

\begin{prop}
The average-case PAR with respect to player $1$ of the trivial and $1$-first protocols for \intersection\ is $1$.  The average-case PAR with respect to player $2$ of the trivial and $1$-first protocols for \intersection is $\left(\frac{3}{2}\right)^k$.
\end{prop}
\begin{proof}
The $1$-partition induced by the trivial protocol is exactly the ideal $1$-partition, from which the first claim follows.

For the second claim, we let $v_k$ be the value of the sum in Eq.~\ref{eq:pargen}.  Let $S$ be a tile in the induced $2$-tiling of the $k$-bit input space; we will also use $S$ to denote the $1$-first-protocol transcript that labels $S$.  We now consider the tiles corresponding to $S$ in the induced $2$-tiling of the $(k+1)$-bit input space.  The tile $10S$ in the bottom-left quadrant is contained in an ideal region that is twice as big as the one that contains $S$---this ideal region contains points in both the bottom-left and top-left quadrants; the same is true of the tile $0S$ in the top-left quadrant.  The tile $0S$ in the top-right quadrant (which is a different $2$-induced tile than the one in the top-left quadrant) is contained in an ideal region that is the same size as the ideal region containing $S$---this ideal region does not contain any points in the bottom-right quadrant.  Finally, the tile $11S$ in the bottom-right quadrant is contained in an ideal region that is the same size as the ideal region containing $S$.  Thus, we have that $v_{k+1} = 6v_k$; by inspection, $v_1=6$, so $v_k = 6^k$.  Note that the average-case PAR with respect to $2$ equals $v_k/4^k$, completing the proof.
\end{proof}

\begin{cor}
The average-case subjective PAR of the trivial and $1$-first protocols for \intersection\ with respect to the uniform distribution is
\[
\left(\frac{3}{2}\right)^k.
\]
\end{cor}

\begin{cor}
If $\PAR^\mathrm{trivial}_i$ denotes the average-case PAR w.r.t. $i$ of the trivial protocol for \intersection\ w.r.t. the uniform distribution, and if $\PAR^\mathrm{1-first}_i$ denotes the average-case PAR w.r.t. $i$ of the $1$-first protocol for \intersection\ w.r.t. the uniform distribution, then
\[
\frac{\PAR^\mathrm{trivial}_2}{\PAR^\mathrm{trivial}_1} = \frac{\PAR^\mathrm{1-first}_2}{\PAR^\mathrm{1-first}_1} = \left(\frac{3}{2}\right)^k.
\]
\end{cor}

\subsubsection{Subjective PAR for the alternating protocol}

\begin{prop}
The average-case PAR with respect to player $1$ of the alternating protocol for \intersection\ is $\frac{4}{5}\left(\frac{5}{4}\right)^k$.  The average-case PAR with respect to player $1$ of the alternating protocol for \intersection\ is $\frac{6}{5}\left(\frac{5}{4}\right)^k$.
\end{prop}
\begin{proof}
We let
\[
h_k = \sum_S |R_1^I(S)|,
\]
where the sum is taken over all induced $1$-rectangles (``horizontal rectangles'') in the $k$-bit value space, and we let
\[
h_k = \sum_S |R_2^I(S)|,
\]
where the sum is taken over all induced $2$-rectangles (``vertical rectangles'') in the $k$-bit value space.

Making use of the structure of the tiling, we have that
\[
v_{k+1} = 2v_k + 2h_k + h_k + v_k = 3(v_k+h_k),
\]
where the summands correspond to the contributions from each quadrant (clockwise from the bottom-left quadrant).  We also have
\[
h_{k+1} = h_k + 2v_k + h_k = 2(v_k+h_k),
\]
where the summands correspond to the contributions from the bottom-left, top-two, and bottom-right quadrants, respectively.  By inspection, we have $h_1 = 4$ and $v_1 = 6$; this gives $h_k = 4\cdot 5^{k-1}$ and $v_k = 6\cdot 5^{k-1}$.  Denoting by $\PAR^\mathrm{alt}_i(k)$ the average-case PAR w.r.t. $i$ of the trivial protocol for \intersection\ w.r.t. the uniform distribution, we have
\begin{eqnarray*}
\PAR^\mathrm{alt}_1(k) &=& \frac{h_k}{4^k} = \frac{4}{5}\left(\frac{5}{4}\right)^k\\
\PAR^\mathrm{alt}_2(k) &=& \frac{v_k}{4^k} = \frac{6}{5}\left(\frac{5}{4}\right)^k
\end{eqnarray*}
as claimed.
\end{proof}

\begin{cor}
The average-case subjective PAR of the alternating protocol for \intersection\ is
\[
\frac{6}{5}\left(\frac{5}{4}\right)^k.
\]
\end{cor}

\begin{cor}
If $\PAR^\mathrm{alt}_i$ denotes the average-case PAR w.r.t. $i$ of the trivial protocol for \intersection\ w.r.t. the uniform distribution, then
\[
\frac{\PAR^\mathrm{alt}_2}{\PAR^\mathrm{alt}_1} = \frac{3}{2}.
\]
\end{cor}

\section{Conclusions and Future Work}\label{sec:conc}

Our definitions of PARs involve the intuitive notion of the
indistinguishability of inputs that is natural to consider in the
context of privacy preservation. Other definitions of PARs may be
appropriate in analyzing other notions of privacy. For example, if
there is a natural notion of ``distance'' between inputs (as
in the examples considered in this paper), one might prefer
protocols that cannot distinguish among a few inputs that are far
from each other to protocols that cannot distinguish among many
inputs that are all relatively close. This necessitates different
definitions of PARs and suggests many interesting avenues for
future work.

Starting from the same place that we did,
namely~\cite{CK91,K92}, Bar-Yehuda {\it et al.}~\cite{BCKO} provided
three definitions of approximate privacy.  We show in~\cite{fjs09tr14} that
the formulation in \cite{BCKO} is not equivalent to ours, but there is more
to do along these lines.  The definition in \cite{BCKO} that seems most
relevant to the study of privacy-approximation ratios is their
notion of {\it h-privacy}. Determine when and how it is possible to
express PARs in terms of $h$-privacy and {\it vice versa}.

Lower bounds on the average-case subjective PARs for \disjoint\ and \intersection\ would be interesting; as noted above, we conjecture that these are exponential in $k$.  Our PAR framework should also be applied to other functions and extended to $n$-party communication.

\section*{Acknowledgements}

We are grateful to audiences at University Residential Centre of Bertinoro, Boston University, DIMACS, the University of Massach\-usetts, Northwestern, Princeton, and Rutgers for helpful questions and feedback.

\appendix

\section{Perfect Privacy and Communication Complexity}\label{ap:model}

For convenience, we include Sec.~2 of (a revised version of) \cite{fjs09tr14} as the text of this appendix.
It contains the basic definitions of communication complexity and privacy that underlie our approach to
approximate privacy.

\subsection{Two-Party Communication Model}\label{subsec-Yao}

We now briefly review Yao's model of two-party communication and
notions of objective and subjective perfect privacy; see Kushilevitz
and Nisan~\cite{KN97} for a comprehensive overview of communication
complexity theory.  Note that we only deal with \emph{deterministic}
communication protocols. Our definitions can be extended to
randomized protocols.

There are two parties, $1$ and $2$, each holding a $k$-bit
\emph{input string}. The input of party $i$, $x_i\in\{0,1\}^k$, is
the \emph{private information} of $i$. The parties communicate with
each other in order to compute the value of a function
$f:\{0,1\}^k\times \{0,1\}^k\rightarrow \{0,1\}^t$. The two parties
alternately send messages to each other. In communication round $j$, one of the parties sends a bit $q_j$ that is a
function of that party's input and the history $(q_1,\ldots,q_{j-1})$ of previously sent messages.  We say that a bit is \emph{meaningful} if it is not a constant function of this input and history and if, for every meaningful bit transmitted previously, there some combination of input and history for which the bit differs from the earlier meaningful bit.  Non-meaningful bits (\emph{e.g.}, those sent as part of protocol-message headers) are irrelevant to our work here and will be ignored.  A \emph{communication
protocol} dictates, for each party, when it is that party's turn to
transmit a message and what message he should transmit, based on the
history of messages and his value.

A communication protocol $P$ is said to compute $f$ if, for every
pair of inputs $(x_1,x_2)$, it holds that $P(x_1,x_2)=f(x_1,x_2)$. As
in~\cite{K92}, the last message sent in a protocol $P$ is assumed to
contain the value $f(x_1,x_2)$ and therefore may require up to $t$
bits. The \emph{communication complexity} of a protocol $P$ is the
maximum, over all input pairs, of the number of bits transmitted during the execution of $P$.

Any function $f:\{0,1\}^k\times \{0,1\}^k\rightarrow\{0,1\}^t$ can
be visualized as a $2^k\times 2^k$ matrix with entries in
$\{0,1\}^t$, in which the rows represent the possible inputs of
party $1$, the columns represent the possible inputs of party $2$,
and each entry contains the value of $f$ associated with its row and
column inputs. This matrix is denoted by $A(f)$.

\begin{definition}[Regions, partitions]
A \emph{region} in a matrix $A$ is any subset of entries in $A$ (not
necessarily a submatrix). A \emph{partition} of $A$ is a collection
of disjoint regions in $A$ whose union equals $A$.
\end{definition}

\begin{definition}[Monochromaticity]
A region $R$ in a matrix $A$ is called \emph{monochromatic} if all
entries in $R$ contain the same value. A \emph{monochromatic
partition} of $A$ is a partition all of whose regions are
monochromatic.
\end{definition}

Of special interest in communication complexity are specific kinds
of regions and partitions called rectangles, and tilings,
respectively:

\begin{definition}[Rectangles, Tilings]\label{def:rect}
A \emph{rectangle} in a matrix $A$ is a submatrix of $A$. A
\emph{tiling} of a matrix $A$ is a partition of $A$ into rectangles.
\end{definition}

\begin{definition}[Refinements]
A tiling $T_1(f)$ of a matrix $A(f)$ is said to be a
\emph{refinement} of another tiling $T_2(f)$ of $A(f)$ if every
rectangle in $T_1(f)$ is contained in some rectangle in $T_2(f)$.
\end{definition}

Monochromatic rectangles and tilings are an important concept in
com\-mu\-ni\-ca\-tion-com\-plex\-i\-ty theory, because they are linked to the
execution of communication protocols. Every communication protocol
$P$ for a function $f$ can be thought of as follows:
\begin{enumerate}
\item Let $R$ and $C$ be the sets of row and column indices of $A(f)$, respectively.  For $R'\subseteq R$ and $C'\subseteq C$, we will abuse notation and write $R'\times C'$ to denote the submatrix of $A(f)$ obtained by deleting the rows not in $R'$ and the columns not in $C'$.

\item While $R\times C$ is not monochromatic:

\begin{itemize}

\item One party $i\in \{0,1\}$ sends a single bit $q$ (whose value
is based on $x_i$ and the history of communication).

\item If $i=1$, $q$ indicates whether $1$'s value is in one of
two disjoint sets $R_1,R_2$ whose union equals $R$. If $x_1\in R_1$,
both parties set $R=R_1$. If $x_1\in R_2$, both parties set $R=R_2$.

\item If $i=2$, $q$ indicates whether $2$'s value is in one of
two disjoint sets $C_1,C_2$ whose union equals $C$. If $x_2\in C_1$,
both parties set $C=C_1$. If $x_2\in C_2$, both parties set $C=C_2$.
\end{itemize}

\item One of the parties sends a last message (consisting of up to $t$ bits) containing the value in all
entries of the monochromatic rectangle $R\times C$.
\end{enumerate}

Observe that, for every pair of private inputs $(x_1,x_2)$, $P$
terminates at some monochromatic rectangle in $A(f)$ that contains
$(x_1,x_2)$. We refer to this rectangle as ``\emph{the monochromatic
rectangle induced by $P$ for $(x_1,x_2)$}''. We refer to the tiling
that consists of all rectangles induced by $P$ (for all pairs of
inputs) as  ``\emph{the monochromatic tiling induced by $P$}''.

\begin{figure}[htp]
\begin{center}
\includegraphics[width=1in]{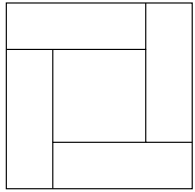}
\caption{\small A tiling that is not induced by any communication
protocol~\cite{K92}}\label{fig:nomech}
\end{center}
\end{figure}

\begin{remark}
There are monochromatic tilings that cannot be induced by communication
protocols. For example, observe that the tiling in
Fig.~\ref{fig:nomech} (which is essentially an example from~\cite{K92}) has this property.
\end{remark}

\subsection{Perfect Privacy}\label{subsec_perfect-privacy}

Informally, we say that a two-party protocol is \emph{perfectly
privacy-preserving} if the two parties (or a third party observing
the communication between them) cannot learn more from the execution
of the protocol than the value of the function the protocol
computes.  (This definition can be extended naturally to protocols
involving more than two participants.)

Formally, let $P$ be a communication protocol for a function $f$.
The \emph{communication string} passed in $P$ is the concatenation
of all the messages $(q_1,q_2, \ldots)$ sent in the course of the
execution of $P$. Let $s_{(x_1,x_2)}$ denote the communication
string passed in $P$ if the inputs of the parties are $(x_1,x_2)$.
We are now ready to define perfect privacy. The following two
definitions handle privacy from the point of view of a party $i$
that does not want the other party (that is, of course, familiar not
only with the communication string, but also with \emph{his own}
value) to learn more than necessary about $i$'s private information.
We say that a protocol is perfectly private with respect to party
$1$ if $1$ never learns more about party $2$'s private information
than necessary to compute the outcome.

\begin{definition} [Perfect privacy with respect to 1]~\cite{CK91,K92} \label{def_ppwrt1}
$P$ is \emph{perfectly private with respect to party $1$} if, for
every $x_2,x'_2$ such that $f(x_1,x_2)=f(x_1,x'_2)$, it holds that
$s_{(x_1,x_2)}=s_{(x_1,x'_2)}$.
\end{definition}

Informally, Def.~\ref{def_ppwrt1} says that party 1's knowledge of the communication
string passed in the protocol and his knowledge of $x_1$ do not
aid him in distinguishing between two possible inputs of $2$.
Similarly:
\begin{definition}[Perfect privacy with respect to 2]~\cite{CK91,K92}
$P$ is \emph{perfectly private with respect to party $2$} if, for
every $x_1,x'_1$ such that $f(x_1,x_2)=f(x'_1,x_2)$, it holds that
$s_{(x_1,x_2)}=s_{(x'_1,x_2)}$.
\end{definition}

\begin{obs}
For any function $f$, the protocol in which party $i$ reveals $x_i$
and the other party computes the outcome of the function is
perfectly private with respect to $i$.
\end{obs}

\begin{definition} [Perfect subjective privacy]
$P$ achieves \emph{perfect subjective privacy} if it is perfectly
private with respect to both parties.
\end{definition}

The following definition considers a different form of privacy---privacy from \emph{a third party} that observes the communication
string but has no \emph{a priori} knowledge about the private
information of the two communicating parties. We refer to this
notion as ``\emph{objective privacy}''.

\begin{definition} [Perfect objective privacy]
$P$ achieves \emph{perfect objective privacy} if, for every two
pairs of inputs $(x_1,x_2)$ and $(x'_1,x'_2)$ such that
$f(x_1,x_2)=f(x'_1,x'_2)$, it holds that
$s_{(x_1,x_2)}=s_{(x'_1,x'_2)}$.
\end{definition}

Kushilevitz~\cite{K92} was the first to point out the interesting
connections between perfect privacy and communication-complexity theory. Intuitively, we can think of any monochromatic
rectangle $R$ in the tiling induced by a protocol $P$ as a set of
inputs that are \emph{indistinguishable} to a third party. This is
because, by definition of $R$, for any two pairs of inputs in $R$, the
communication string passed in $P$ must be the same. Hence we can
think of the privacy of the protocol in terms of the tiling induced
by that protocol.

Ideally, every two pairs of inputs that are assigned the same
outcome by a function $f$ will belong to the same monochromatic
rectangle in the tiling induced by a protocol for $f$. This
observation enables a simple characterization of perfect
privacy-preserving mechanisms.

\begin{definition} [Ideal monochromatic partitions]
A monochromatic region in a matrix $A$ is said to be a {\em maximal
monochromatic region} if no monochromatic region in $A$ properly
contains it. \emph{The ideal monochromatic partition} of $A$ is made
up of the maximal monochromatic regions.
\end{definition}

\begin{obs}
For every possible value in a matrix $A$, the maximal monochromatic
region that corresponds to this value is unique. This implies the
uniqueness of the ideal monochromatic partition for $A$.
\end{obs}

\begin{obs}[A characterization of perfectly privacy-preserving protocols]\label{obs_privacy=maximal}
\ \\ A communication protocol $P$ for $f$ is perfectly privacy-preserving
iff the monochromatic til\-ing induced by $P$ is the ideal
monochromatic partition of $A(f)$. This holds for all of the above
notions of privacy.
\end{obs}

\section{Other Notions of Approximate Privacy}\label{ap:alternate}

For the convenience of the reader, we repeat the discussion from Sec.~6.1 of~\cite{fjs09tr14} (the revision dated the same date as this report) of other possible approaches to approximate privacy.

By our definitions, the worst-case/average-case PARs of a protocol
are determined by the worst-case/expected value of the expression
$\frac{|R^I(\mathbf{x})|}{|R^P(\mathbf{x})|}$, where $R^P(\mathbf{x})$ is the
monochromatic rectangle induced by $P$ for input $\mathbf{x}$, and
$R^I(\mathbf{x})$ is the monochromatic region containing
$A(f)_{\mathbf{x}}$ in the ideal monochromatic partition of $A(f)$.
That is, informally, we are interested in the ratio of the
\emph{size} of the ideal monochromatic region for a specific pair of inputs to the \emph{size} of the monochromatic rectangle induced by the protocol for that pair. More generally, we can define worst-case/average-case PARs with respect to a function $g$ by considering the ratio
$\frac{g(R^I(\mathbf{x}),\mathbf{x})}{g(R^P(\mathbf{x}),\mathbf{x})}$.  Our definitions of PARs set $g(R,\mathbf{x})$ to be the cardinality of $R$.  This captures the intuitive notion of the
indistinguishability of inputs that is natural to consider in the
context of privacy preservation. Other definitions of PARs may be
appropriate in analyzing other notions of privacy.  We suggest a few here; further
investigation of these and other definitions provides many interesting avenues for future work.

\begin{figure}[htp]
\begin{center}
\includegraphics[width=4in]{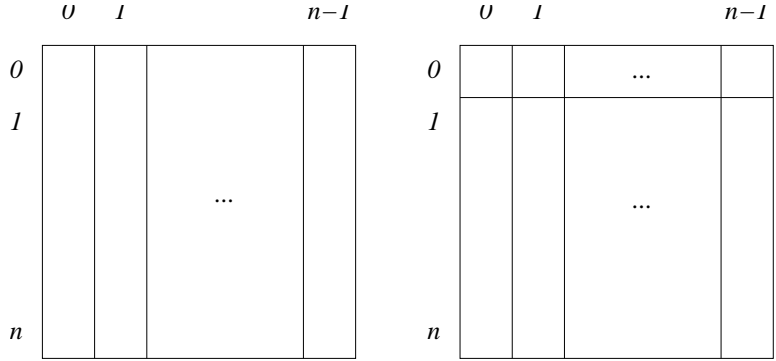}
\caption{\small Maximal monochromatic regions (left) and protocol-induced rectangles (right) for an example showing the deficiencies of PAR definitions based on probability mass. }\label{fig:prob-mass}
\end{center}
\end{figure}

\textbf{Probability mass.}\ \ Given a probability distribution $D$ over the parties' inputs, a
seemingly natural choice of $g$ is the probability mass. That is,
for any region $R$, $g(R)=Pr_D(R)$, the probability (according to
$D$) that the input corresponds to an entry in $R$. However, a 
simple example illustrates that this intuitive choice of
$g$ is problematic: Consider a problem for which $\{0,\ldots,n\}\times\{i\}$
is a maximal monochromatic region for $0\leq i\leq n-1$ as illustrated in the left part of Fig.~\ref{fig:prob-mass}.  Let $P$ be the communication protocol consisting
of a single round in which party $1$ reveals whether or not his value is $0$;
this induces the monochromatic tiling with tiles $\{(0,i)\}$ and $\{(1,i),\ldots,(n,i)\}$ for each $i$ as illustrated in the right part of Fig.~\ref{fig:prob-mass}.  Now, let $D_1$ and $D_2$ be the probability distributions over the inputs $\mathbf{x}=(x_1,x_2)$ such that, for $0\leq i\leq n-1$ and $1\leq j\leq n$, $Pr_{D_1}[(x_1,x_2)=(0,i)]=\frac{\epsilon}{n}$, $Pr_{D_1}[(x_1,x_2)=(j,i)]=\frac{1-\epsilon}{n^2}$, $Pr_{D_2}[(x_1,x_2)=(0,i)]=\frac{1-\epsilon}{n}$, and $Pr_{D_2}[(x_1,x_2)=(j,i)]=\frac{\epsilon}{n^2}$ for some small $\epsilon>0$.  
Intuitively, any
reasonable definition of PAR should imply that, for $D_1$, $P$
provides ``bad'' privacy guarantees (because w.h.p.~it reveals the 
value of $x_1$), 
and, for $D_2$, $P$ provides ``good''
privacy (because w.h.p.~it reveals little 
about $x_1$). 
In sharp contrast, choosing $g$ to be the probability mass results
in the same average-case PAR in both cases.

\textbf{Other additive functions.}\ \ In our definition of PAR and in the probability-mass approach, each input $\mathbf{x}$ in a rectangle contributes to $g(R,\mathbf{x})$ in a way that is independent of the other inputs in $R$.  Below, we discuss some natural approaches that violate this condition, but we start by noting that other functions that satisfy this condition may be of interest.  For example, taking $g(R,\mathbf{x}) = 1+\sum_{\mathbf{y}\in R\setminus \mathbf{x}} d(\mathbf{x},\mathbf{y})$, where $d$ is some distance defined on the input space, gives our original definition of PAR when $d(x,y) = 1-\delta_{\mathbf{x},\mathbf{y}}$ and might capture other interesting definitions (in which indistinguishable inputs that are farther away from $\mathbf{x}$ contribute more to the privacy for $\mathbf{x}$).  (The addition of $1$ ensures that the ratio $g(R^I,\mathbf{x})/g(R^P,\mathbf{x})$ is defined, but that can be accomplished in other ways if needed.)\ \ Importantly, here and below, the notion of distance that is used might not be a Euclidean metric on the $n$-player input space $[0,2^k-1]^n$.  It could instead (and likely would) focus on the problem-specific interpretation of the input space.  Of course, there are may possible variations on this (\eg, also accounting for the probability mass).

\textbf{Maximum distance.}\ \ We might take the view that a protocol does not reveal much about an input $\mathbf{x}$ if there is another input that is ``very different'' from $\mathbf{x}$ that the protocol cannot distinguish from $\mathbf{x}$ (even if the total number of things that are indistinguishable from $\mathbf{x}$ under the protocol is relatively small).  For some distance $d$ on the input space, we might than take $g$ to be something like $1+\max_{\mathbf{y}\in R\setminus\{\mathbf{x}\}} d(\mathbf{y},\mathbf{x})$.

\textbf{Plausible deniability.}\ \ One drawback to the maximum-distance approach is that it does not account for the probability associated with inputs that are far from $\mathbf{x}$ (according to a distance $d$) and that are indistinguishable from $\mathbf{x}$ under the protocol.  While there might be an input $\mathbf{y}$ that is far away from $\mathbf{x}$ and indistinguishable from $\mathbf{x}$, the probability of $\mathbf{y}$ might be so small that the observer feels comfortable assuming that $\mathbf{y}$ does not occur.  A more realistic approach might be one of ``plausible deniability.''\ \ This makes use of a plausibility threshold---intuitively, the minimum probability that the ``far away'' inputs(s) (which is/are indistinguishable from $\mathbf{x}$) must be assigned in order to ``distract'' the observer from the true input $\mathbf{x}$.  This threshold might correspond to, \eg, ``reasonable doubt'' or other levels of certainty.  We then consider how far we can move away from $\mathbf{x}$ while still having ``enough'' mass (\ie, more than the plausibility threshold) associated with the elements indistinguishable from $\mathbf{x}$ that are still farther away.  We could then take $g$ to be something like $1+\max \{d_0 | Pr_D(\{\mathbf{y}\in R|d(\mathbf{y},\mathbf{x})\geq d_0\})/Pr_D(R) \geq t\}$; other variations might focus on mass that is concentrated in a particular direction from $\mathbf{x}$.  (In quantifying privacy, we would expect to only consider those $R$ with positive probability, in which case dividing by $Pr_D(R)$ would not be problematic.)\ \ Here we use $Pr_D(R)$ to normalize the weight that is far away from $x$ before comparing it to the threshold $t$; intuitively, an observer would know that the value is in the same region as $x$, and so this seems to make the most sense.

\textbf{Relative rectangle size.}\ \ One observation is that a bidder likely has a very different view of an auctioneer's being able to tell (when some particular protocol is used) whether his bid lies between $995$ and $1005$ than he does of the auctioneer's being able to tell whether his bid lies between $5$ and $15$.  In each case, however, the bids in the relevant range are indistinguishable under the protocol from $11$ possible bids.  In particular, the privacy gained from an input's being distinguishable from a fixed number of other inputs may (or may not) depend on the context of the problem and the intended interpretation of the values in the input space.  This might lead to a choice of $g$ such as $diam_d(R)/|\mathbf{x}|$, where $diam_d$ is the diameter of $R$ with respect to some distance $d$ and $|\mathbf{x}|$ is some (problem-specific) measure of the size of $\mathbf{x}$ (\eg, bid value in an auction).  Numerous variations on this are natural and may be worth investigating.

\textbf{Information-theoretic approaches.}\ \ Information-the\-oretic approaches using conditional entropy are also natural to consider when studying privacy, and these have been used in various settings.  Most relevantly, Bar-Yehuda \etal.~\cite{BCKO} defined multiple measures based on the conditional mutual information about one player's value (viewed as a random variable) revealed by the protocol trace and knowledge of the other player's value.  It would also be natural to study objective-PAR versions using the entropy of the random variable corresponding to the (multi-player) input conditioned only on the protocol output (and not the input of any player).  Such approaches might facilitate the comparison of privacy between different problems.

\end{document}